\documentclass[pdflatex,tex,twocolumn,epjc3]{svjour3}

\usepackage{amsmath}
\usepackage{amssymb}
\usepackage{amsfonts}
\usepackage{mathrsfs}
\usepackage{bm}      
\usepackage{slashed}
\usepackage{latexsym}
\usepackage{enumitem}

\usepackage[T1]{fontenc}
\usepackage{lmodern}
\usepackage{anyfontsize}

\usepackage{color}
\usepackage{hyphenat}
\usepackage{float}
\usepackage{graphicx}
\usepackage{indentfirst}
\usepackage{setspace}
\usepackage{ragged2e}

\usepackage{booktabs}
\usepackage{multirow}
\usepackage{tabularx}
\usepackage{makecell}
\usepackage{dcolumn}
\usepackage{adjustbox}

\usepackage{microtype}
\usepackage[numbers,sort&compress]{natbib}

\usepackage[colorlinks,
            linkcolor=blue,
            citecolor=blue,
            urlcolor=blue,
            breaklinks=true]{hyperref}

\journalname{Eur. Phys. J. C}

\begin{document}

\title{Cosmological Constraints on the DGP Model in light of DESI DR2 2025 Data}

\author{Xinyi Dai\thanksref{addr1}, Yupeng Yang\thanksref{addr1,e1} and Yicheng Wang\thanksref{addr1}}
\thankstext{e1}{e-mail: ypyang@aliyun.com}

\institute{School of Physics and Physical Engineering, Qufu Normal University, Qufu, Shandong, 273165, China \label{addr1}}
\date{Received: date / Accepted: date}

\maketitle
\begin{abstract}
We present updated constraints on both flat and non-flat Dvali–Gabadadze–Porrati (DGP) cosmological models using the latest baryon acoustic oscillation (BAO) measurements from the Dark Energy Spectroscopic Instrument Data Release 2 (DESI DR2), in combination with cosmic chronometer (CC), Type Ia supernova (SNIa), and cosmic microwave background (CMB) distance priors. 
For the non-flat DGP model, we obtain $H_0 = 64.05 \pm 0.27\, \mathrm{km\,s^{-1}\,Mpc^{-1}}$, $\Omega_m = 0.3264 \pm 0.0043$, and $\Omega_k = 0.0088 \pm 0.0016$, corresponding to a transition redshift $z_t \simeq 0.41$. 
For the flat case, the constraints are $H_0 = 63.28 \pm 0.25\, \mathrm{km\,s^{-1}\,Mpc^{-1}}$ and $\Omega_m = 0.3303 \pm 0.0036$. 
In both scenarios, the inferred Hubble constant is significantly lower than the \textit{Planck} $\mathrm{\Lambda}$CDM value, indicating that the DGP framework does not alleviate the Hubble tension. 
Current observations strongly disfavor the DGP framework, primarily due to its inability to simultaneously accommodate DESI BAO and CMB constraints.
By incorporating the latest high-precision DESI observations within a unified analysis framework, this work provides updated and more stringent limits on the DGP scenario, offering a consolidated assessment of its viability in the context of current cosmological data.
\end{abstract}

\section{Introduction}
The late-time acceleration of the Universe, initially revealed by the Type Ia supernovae observations~\cite{SupernovaCosmologyProject:1998vns,SupernovaCosmologyProject:1997zqe,SupernovaSearchTeam:1998fmf}, challenges the expectation of a decelerating, matter-dominated expansion. This phenomenon has since been robustly corroborated by a suite of independent cosmological probes, including the cosmic microwave background (CMB) measurements~\cite{Caldwell:2003hz,Huang:2005re}, baryon acoustic oscillations (BAO)~\cite{BOSS:2016wmc}, and cosmic chronometers (CC)~\cite{Jimenez:2001gg,Pinho:2018unz}. Despite this observational consensus, the physical mechanism driving this acceleration remains one of the most profound puzzles in modern cosmology.

Within the standard $\mathrm{\Lambda}$CDM framework, the acceleration is attributed to the cosmological constant, $\mathrm{\Lambda}$. However, the observed value of $\mathrm{\Lambda}$ is approximately 121 orders of magnitude smaller than predictions from quantum field theory~\cite{Copeland:2006wr}, a discrepancy known as the cosmological constant problem or the fine-tuning problem. Furthermore, the persistent $\sim 5\sigma$ tension in the Hubble constant between early-Universe constraints from Planck ($H_0 = 67.4 \pm 0.54\,\mathrm{km\,s^{-1}\,Mpc^{-1}}$)~\cite{Planck:2018vyg} and local distance ladder determinations ($H_0 = 73.04 \pm 1.04\,\mathrm{km\,s^{-1}\,Mpc^{-1}}$)~\cite{Riess:2021jrx} has intensified the search for alternative frameworks.
These challenges suggest that the late-time acceleration may arise either from an exotic dark energy component or from modifications to general relativity on large scales. 
In the former case, a variety of dynamical dark energy models, such as quintessence, k-essence, and interacting dark energy scenarios, have been proposed as possible resolutions to the Hubble tension and the cosmological constant problem~\cite{Guo:2018ans,Li:2025vuh,Li:2010xjz,Xu:2016grp,Dai:2026pvx,Wang:2026rkx,Yang:2025vnm,Hu:2023jqc,Yang:2025oax,Li:2025dwz,Liu:2018kjv,Li:2025owk,Li:2026xaz,Li:2026asg}.

To address these conceptual and observational challenges, a wide array of modified gravity theories has been developed by generalizing the Einstein-Hilbert action. A prominent class is $f(R)$ gravity~\cite{Nojiri:2010wj}, which replaces the Ricci scalar $R$ (the standard measure of spacetime curvature) with an arbitrary functional form. Further extensions, such as $f(R,T)$ and $f(R,L_m)$ gravity~\cite{2022IJGMM..1950038K,Koussour:2022xcv,Harko:2014axa,Koussour:2026uyi,2022IJGMM..1950038K}, introduce explicit non-minimal couplings between the geometry and the matter sector, where $T$ is the trace of the energy-momentum tensor and $L_m$ is the matter Lagrangian density. Alternatively, the symmetric teleparallel framework offers a distinct geometric perspective by attributing gravity to the non-metricity scalar $Q$ rather than curvature, leading to models such as $f(Q)$ and $f(Q,T)$ theories~\cite{Myrzakulov:2023qjo,Koussour:2022nsc,Myrzakulov:2023sir,Koussour:2025yus,Lin:2021uqa,Frusciante:2021sio}.
The observational viability of these diverse frameworks has been extensively tested within the FLRW cosmology using various cosmic probes, such as CC, BAO, and Type Ia supernovae, to provide robust constraints on their respective parameter spaces~\cite{Koussour:2024kxd,Koussour:2023rmm}.

A conceptually distinct approach is provided by the Dvali-Gabadadze-Porrati (DGP) braneworld model~\cite{Dvali:2000xg,Lue:2005ya,Charmousis:2006pn,Sahni:2002dx}, which posits that our four-dimensional (4D) Universe is a brane embedded in a five-dimensional (5D) Minkowski bulk. In this paradigm, gravity is characterized by a crossover scale $r_c = M_{\mathrm{Pl}}^2/(2M_5^3)$, which marks the transition between the 4D and higher-dimensional gravitational regimes. On scales much smaller than $r_c$, gravity is effectively trapped on the brane, recovering the standard $1/r^2$ behavior of General Relativity. Conversely, on cosmological scales comparable to $r_c$, gravity begins to "leak" into the extra dimension, which naturally drives an effective late-time acceleration without the need for a fine-tuned cosmological constant. This mechanism offers a testable geometric alternative through precision measurements of the expansion history $H(z)$ and its sensitivity to the spatial curvature $\Omega_k$.

Historically, the DGP model has been subjected to rigorous testing across a diverse array of cosmological probes. These include distance measurements from Type Ia Supernovae (SNIa)~\cite{Zhu:2004vj,Guo:2006ce,Liang_2011,Movahed:2007ie,Barger:2006vc,Xu:2010um,Deffayet:2002sp,Liu:2018kjv,Xu:2016grp}, global constraints from CMB, and structural traces from BAO and CC~\cite{Guo:2006ce,Liang_2011,Movahed:2007ie,Xu:2010um,PhysRevD.84.023005,Rydbeck:2007gy,Liu:2018kjv,Lombriser:2009xg,Mishra:2025goj}. Complementary limits have also been derived from gravitational lensing~\cite{PhysRevD.84.023005,Zhu:2008sg,Cao_2012,Liu:2019ddm}, gamma-ray bursts (GRBs)~\cite{Liang_2011,Xu:2010um}, and galaxy cluster observations~\cite{Liang_2011,Movahed:2007ie,Xu:2010um}. While these collective studies established the general viability of the DGP framework as a geometric alternative to dark energy, their constraining power was inevitably hindered by the statistical uncertainties and limited redshift coverage of earlier observational catalogs.

The landscape of observational cosmology has been recently transformed by the Dark Energy Spectroscopic Instrument (DESI). 
The BAO measurements from the DESI Data Release 2 (DR2)~\cite{DESI:2025zgx}—incorporating three years of observations with over 14 million galaxies and quasars extending up to redshift $z \approx 4.2$—offer unprecedented statistical power and redshift resolution.
This dataset provides among the most stringent constraints to date on the cosmic expansion history, presenting a unique opportunity to probe the viable parameter space of modified gravity. Notably, recent analyses of DESI DR2 have already reported intriguing deviations from the $\mathrm{\Lambda}$CDM paradigm, with some studies favoring specific $f(R)$ gravity scenarios~\cite{Plaza:2025gcv,Escamilla:2024xmz,Chudaykin:2024gol} over a cosmological constant. Such developments provide a compelling motivation to re-evaluate the DGP framework using this high-precision observational baseline.

In this work, we perform a systematic joint likelihood analysis of both flat and non-flat DGP models. Our study leverages the DESI DR2 BAO datasets, complemented by the Pantheon SNIa sample, CC, and CMB distance priors from Planck 2018. By focusing on this robust data combination, we aim to determine whether the DGP framework can provide a consistent description of the Universe's evolution in the era of precision cosmology.
The structure of this paper is organized as follows. Section~\ref{sec:2} summarizes the theoretical foundations and key cosmological features of the DGP models under investigation. Section~\ref{sec:3} describes the observational datasets and the statistical methodology employed for our analysis. 
In Section~\ref{sec:4}, we present the updated constraints and a comprehensive analysis of the DGP model, including consistency tests, dataset combinations, and dynamical evolution, with comparisons to previous results.
Finally, comprehensive discussions and concluding remarks are presented in Section~\ref{sec:5}.

\section{The DGP model}
\label{sec:2}

In the present paper, we consider a 4D Friedmann-Lema\^{i}tre-Robertson-Walker (FLRW) universe, where the metric is given by~\cite{Deffayet:2000uy}:
\begin{equation}
ds^2 = -dt^2 + a^2(t)\left[ dr^2 + S^2(r)\,d\psi^2 \right],
\end{equation}
where $a(t)$ is the time-dependent 4D scale factor, $\psi$ are angular coordinates, and $S(r) = \sinh(r)$, $S(r) = r$, $S(r) = \sin(r)$ correspond to open, flat, and closed geometries, respectively.
Based on the dynamics of this metric, the standard first Friedmann equation in our model~\cite{Deffayet:2002sp,Deffayet:2001pu} takes the form:
\begin{equation}
H^2 + \frac{k}{a^2} = \left(\sqrt{\frac{\rho}{3M_{\rm{Pl}}^2} + \frac{1}{4r_c^2}} + \frac{1}{2r_c}\right)^2,
\end{equation}
where $\rho$ is the energy density, $M_{\rm{Pl}}$ denotes the 4D Planck mass, and $r_c = M_{\mathrm{Pl}}^2/(2M_5^3)$ is the crossover scale governing the transition between 4D and 5D dynamics, where $M_{\rm{5}}$ denotes the 5D Planck mass.
Assuming a non-flat universe with matter and radiation components, incorporating extra dimensions, the modified Friedmann equation can be expressed as~\cite{Guo:2006ce}:
\begin{equation}
\begin{split}
H^2 = H_0^2 \Biggl[\, &\Omega_k(1 + z)^2 + \biggl( \sqrt{\Omega_{r_c}} \\
&+ \sqrt{\Omega_{r_c} + \Omega_m(1 + z)^3 + \Omega_r(1 + z)^4 }\biggr) \biggr],
\end{split}
\end{equation}
here, $\Omega_k$, $\Omega_m$, and $\Omega_r$ represent the fractional contributions of curvature, matter, and radiation, respectively, while $\Omega_{r_c}$ is defined as $\Omega_{r_c} \equiv 1/(4r_c^2H_0^2)$, denoting the bulk-induced term. 
Here we adopt the standard convention in which $\Omega_k > 0$ corresponds to an open universe, $\Omega_k < 0$ corresponds to a closed universe, and $\Omega_k = 0$ represents a spatially flat universe.
In the non-flat case, the normalization condition is expressed accordingly:
\begin{equation}
\Omega_k + \left( \sqrt{\Omega_{r_c}} + \sqrt{\Omega_{r_c} + \Omega_{m} + \Omega_r} \right)^2 = 1,
\end{equation}
where the flat case corresponds to $\Omega_k = 0$.


To characterize the expansion history of the Universe, we introduce the deceleration parameter $q \equiv -a\ddot{a}/\dot{a}^2$, which can be expressed in terms of the dimensionless Hubble parameter as $q(z) = -1 + \frac{1+z}{2E^2(z)} \frac{dE^2(z)}{dz}$~\cite{Koussour:2024nhw}.
For the non-flat DGP model, by substituting the Friedmann equation into this geometric definition and setting $z=0$, the present-day deceleration parameter $q_0$ is derived as~\cite{Guo:2006ce,Zhu:2003wv}:
\begin{equation}
\begin{split}
q_0 = & \left( \frac{\frac{3}{2}\Omega_m + 2\Omega_r}{\sqrt{\Omega_m + \Omega_r + \Omega_{r_c}}} \right) \\
& \times \left( \sqrt{\Omega_m + \Omega_r + \Omega_{r_c}} + \sqrt{\Omega_{r_c}} \right) + \Omega_k - 1,
\end{split}
\end{equation}
where $\Omega_k$ indirectly influences the deceleration through the constraint equation for the characteristic scale $\Omega_{r_c}$, given by $\sqrt{\Omega_{r_c}} = (1 - \Omega_k - \Omega_m - \Omega_r) / 2\sqrt{1 - \Omega_k}$.
This generalized relation consistently recovers the standard flat DGP expression when $\Omega_k = 0$.

A positive value of $q$ indicates decelerating expansion, while $q < 0$ corresponds to accelerated expansion, with $q = 0$ representing uniform expansion. The epoch at which the Universe transitions from deceleration to acceleration is defined as the transition redshift $z_t$, corresponding to the condition $q(z_t) = 0$, where a larger (smaller) value of $z_t$ indicates an earlier (later) onset of cosmic acceleration.
For the flat DGP model ($\Omega_k = 0$), this transition redshift can be analytically expressed as~\cite{Zhu:2004vj,Avelino:2001qh,Turner:2001mx}:
\begin{equation}
z_t = -1 + 2\left(\frac{\Omega_{r_c}}{\Omega_m}\right)^{1/3}.
\end{equation}
However, in the more general non-flat case, $z_t$ does not possess a simple closed-form analytical solution due to the contribution of spatial curvature. In such scenarios, $z_t$ is determined by numerically solving the root of $q(z, \Omega_k) = 0$. As a crucial observable, the transition redshift $z_t$ can be effectively constrained by observational data. In the following chapter, we will explore various observational datasets and statistical methods to test this model and constrain the relevant cosmological parameters.

\section{The datasets and methodology}
\label{sec:3}
\subsection{Baryon Acoustic Oscillations}

The Baryon Acoustic Oscillation (BAO) measurements from the DESI Data Release 2 (DR2)~\cite{DESI:2025zgx} provide a powerful geometric probe, utilizing a sample of over 14 million tracers. These measurements demonstrate high statistical robustness, with systematic uncertainties constrained to less than $9\%$ of the total error budget.

The BAO feature is characterized by the sound horizon at the drag epoch, $r_d \equiv r_s(z_d)$, where the sound horizon scale $r_s(z)$ is given by the standard integral:
\begin{equation}
r_s(z) = \int_{z}^{\infty} \frac{c_s(z')}{H(z')} \mathrm{d}z'.
\label{eq:rs}
\end{equation}
Here, the sound speed in the photon-baryon fluid is $c_s(z) = c / \sqrt{3(1 + \bar{R}_b/(1+z))}$. In this work, rather than adopting a fixed $r_d$ value, we compute it dynamically using Eq.~\eqref{eq:rs} to ensure theoretical consistency with the CMB sector~\cite{Eisenstein:1997ik,2010JCAP...07..022H}.

Our analysis utilizes the transverse distance $D_M/r_d$, the Hubble distance $D_H/r_d$, and the isotropic volume distance $D_V/r_d$ from the DESI DR2 catalog. The $\chi^2$ function for the BAO sector is defined as:
\begin{equation}
\chi^2_{\text{DESI}} = \sum_{i} \frac{\left(D_{V,i}^{\text{th}} - D_{V,i}^{\text{obs}}\right)^2}{\sigma_{V,i}^2} + \sum_{j} \Delta D_j^{T} C_{\text{DESI}}^{-1} \Delta D_j,
\end{equation}
where the first term corresponds to the $D_V/r_d$ measurements, and the second term accounts for the correlated $D_M/r_d$ and $D_H/r_d$ measurements across different redshift bins, as dictated by the DR2 inverse covariance matrix $C_{\text{DESI}}^{-1}$.

\subsection{Cosmic Chronometer}
The Hubble rate can be determined using the Cosmic Chronometer (CC) method, which provides a direct measurement of the cosmic expansion rate by estimating the age difference between passively evolving galaxies at different redshifts.
Using the differential age technique~\cite{Jimenez:2001gg}, the Hubble parameter H(z) is derived from the relation:
\begin{equation}
H(z)=-\frac{1}{1+z}\frac{\Delta z}{\Delta t}.
\end{equation}

For this study, we utilize a dataset of 32 H(z)~\cite{Moresco:2022phi} measurements spanning the redshift range 0.07 < z < 1.965. The $\chi^2$ statistic for the CC data is given by:
\begin{equation}
\chi^2_{\text{CC}}=\sum_{i = 1}^{32}\frac{\left[H^{\text{th}}(z_i)-H^{\text{obs}}(z_i)\right]^2}{\sigma_i^2},
\end{equation}
where $H^{\text{obs}}(z_i)$ represents the observed Hubble parameter at redshift $z_i$, and $H^{\text{th}}(z_i)$ denotes the theoretical prediction.

\subsection{Type Ia supernova}
\label{sec:sn}

By virtue of their standardized candle properties, Type Ia supernovae (SNIa) allow for the determination of luminosity distances. The methodology for our statistical analysis follows that described in Ref.~\cite{Yang:2025boq}.

For a general cosmological model with possible non-zero spatial curvature $\Omega_k$, the dimensionless luminosity distance is defined as
\begin{equation}
D_L(z) = \frac{1+z}{\sqrt{|\Omega_k|}} \, \mathrm{sinn} \left[ \sqrt{|\Omega_k|} \int_0^z \frac{\mathrm{d}z'}{E(z', \bm{\theta})} \right],
\label{eq:DL_general}
\end{equation}
where $E(z) = H(z)/H_0$ is the dimensionless Hubble parameter. The function $\mathrm{sinn}(x)$ is defined as $\sin(x)$ for $\Omega_k < 0$, $x$ for $\Omega_k = 0$, and $\sinh(x)$ for $\Omega_k > 0$.
The theoretical distance modulus is expressed as
\begin{equation}
\mu_{\mathrm{th}}(z) = 5\log_{10} D_L(z) + \mu_0,
\end{equation}
where $\mu_0$ is a nuisance parameter that encapsulates the dependence on the Hubble constant $H_0$. This formulation is consistent with Ref.~\cite{Yang:2025boq}.

We use the Pantheon dataset sample~\cite{Pan-STARRS1:2017jku}, consisting of 1048 spectroscopically confirmed SNIa spanning the redshift range $0.01 < z < 2.3$.   
To ensure robust statistical inference, we incorporate the full covariance matrix $\mathbf{C}$, which accounts for both statistical and systematic uncertainties. 

Following the standard methodology for uncalibrated standard candles~\cite{Gong:2007se,PhysRevD.72.123519}, 
we eliminate the dependence on the nuisance parameter $\mu_0$ by adopting the analytical marginalization method. The marginalized $\chi^2_{\mathrm{SN}}$ is constructed as
\begin{equation}
\chi^2_{\text{SNIa}} = A - \frac{B^2}{C},
\end{equation}
where
\begin{equation}
\begin{aligned}
A &= \Delta^{T} \mathbf{C}^{-1} \Delta, \\
B &= \Delta^{T} \mathbf{C}^{-1} \mathbf{1}, \\
C &= \mathbf{1}^{T} \mathbf{C}^{-1} \mathbf{1},
\end{aligned}
\end{equation}
with $\mathbf{1}$ denoting a unit column vector and the residual vector defined as $\Delta_i = \mu_{\mathrm{obs}}(z_i) - \mu_{\mathrm{th}}(z_i).$
This formulation removes the explicit dependence on the nuisance parameter $\mu_0$, thereby avoiding the degeneracy between $\mu_0$ and the Hubble constant $H_0$, and ensures that the SNIa data constrain only the relative distance--redshift relation.

\subsection{CMB distance priors}
To maintain computational efficiency while preserving most of the constraining power of the full CMB measurements, we adopt the distance priors derived from the final \textit{Planck} 2018 data release~\cite{Chen:2018dbv,Planck:2018vyg}. 
This compressed likelihood approach effectively summarizes the CMB information in terms of three parameters: the shift parameter $R$, the acoustic scale $l_A$, and the physical baryon density $\omega_b = \Omega_b h^2$.

The shift parameter $R$ and the acoustic scale $l_A$, which characterize the positions of the acoustic peaks, are defined as~\cite{2020A&A...641A...6P,Xu:2016grp}
\begin{equation}
R = \frac{\sqrt{\Omega_m H_0^2}}{c} (1 + z_\star) D_{\rm A}(z_\star),
\end{equation}
\begin{equation}
l_{\rm A} = \pi (1 + z_\star) \frac{D_{\rm A}(z_\star)}{r_s(z_\star)},
\end{equation}
where $z_\star$ denotes the redshift at the photon decoupling epoch. 
Following Ref.~\cite{Hu:1995en}, $z_\star$ is computed using the fitting formula
\begin{equation}
z_\star = 1048 \left[ 1 + 0.00124(\Omega_b h^2)^{-0.738} \right] 
\left[ 1 + g_1 (\Omega_m h^2)^{g_2} \right],
\end{equation}
with
\begin{equation}
g_1 = \frac{0.0783 (\Omega_b h^2)^{-0.238}}{1 + 39.5 (\Omega_b h^2)^{0.763}}, 
\quad 
g_2 = \frac{0.560}{1 + 21.1 (\Omega_b h^2)^{1.81}}.
\end{equation}
The comoving sound horizon $r_s(z_\star)$ is calculated using the standard integral definition given in Eq.~\eqref{eq:rs}.

The parameter vector is defined as $\mathbf{x} = (R, l_A, \omega_b)$, with observational values
$\mathbf{x}_{\rm obs} = (R^{\rm Pl}, l_A^{\rm Pl}, \omega_b^{\rm Pl})$. 
The corresponding $\chi^2$ is constructed as
\begin{equation}
\chi^2_{\rm CMB} = \Delta \mathbf{x} \, C_{\rm CMB}^{-1} \, \Delta \mathbf{x}^{T},
\end{equation}
where $\Delta \mathbf{x} = \mathbf{x} - \mathbf{x}_{\rm obs}$ and $C_{\rm CMB}$ is the covariance matrix given in Ref.~\cite{Chen:2018dbv}.

\subsection{Constraints on the models}

To ensure the physical consistency of our joint analysis, the sound horizon at the drag epoch, $r_d$, is treated self-consistently rather than as a fixed scale. Specifically, $r_d$ is computed from the standard integral definition $r_d = r_s(z_d) = \int_{z_d}^{\infty} \frac{c_s(z)}{H(z)} \, \mathrm{d}z$, where $c_s(z)$ is the sound speed in the baryon-photon fluid. 
By allowing $r_d$ to vary with the early-time parameters constrained by the \textit{Planck} distance priors, we maintain a robust calibration between the BAO and CMB sectors within the DGP framework. This treatment is crucial for avoiding potential biases in the inferred $H_0$ and $\Omega_k$ that often arise from a fixed-$r_d$ approximation~\cite{Eisenstein:1997ik,2010JCAP...07..022H}.

Building on this consistent framework, we perform a Monte Carlo Markov Chain (MCMC) analysis for the DGP model, adopting uniform priors on the model parameters: 
$\Omega_{m} \in (0, 1)$, $\Omega_{r_c} \in (0, 1)$, $H_0 \in (50, 100)$, $\Omega_k \in (-1, 1)$, and $\Omega_b h^2 \in (0.001, 0.1)$. 
The parameter estimation and visualization are carried out using the GetDist package~\cite{Lewis:2019xzd}.
The best-fit parameters are obtained by maximizing the likelihood function
\begin{equation}
\mathcal{L} \propto \exp\left(-\chi^2_{\rm total}/2\right).
\end{equation}

In this work, we combine four observational datasets, namely DESI DR2 BAO, SNIa, CC, and CMB distance priors. The total $\chi^2$ is given by
\begin{equation}
\chi^2_{\rm total} = \chi^2_{\rm DESI} + \chi^2_{\rm CC} + \chi^2_{\rm SNIa} + \chi^2_{\rm CMB},
\end{equation}
which defines the joint likelihood used in the MCMC analysis.


\begin{figure}[bp]
    \centering
    \includegraphics[width=\columnwidth]{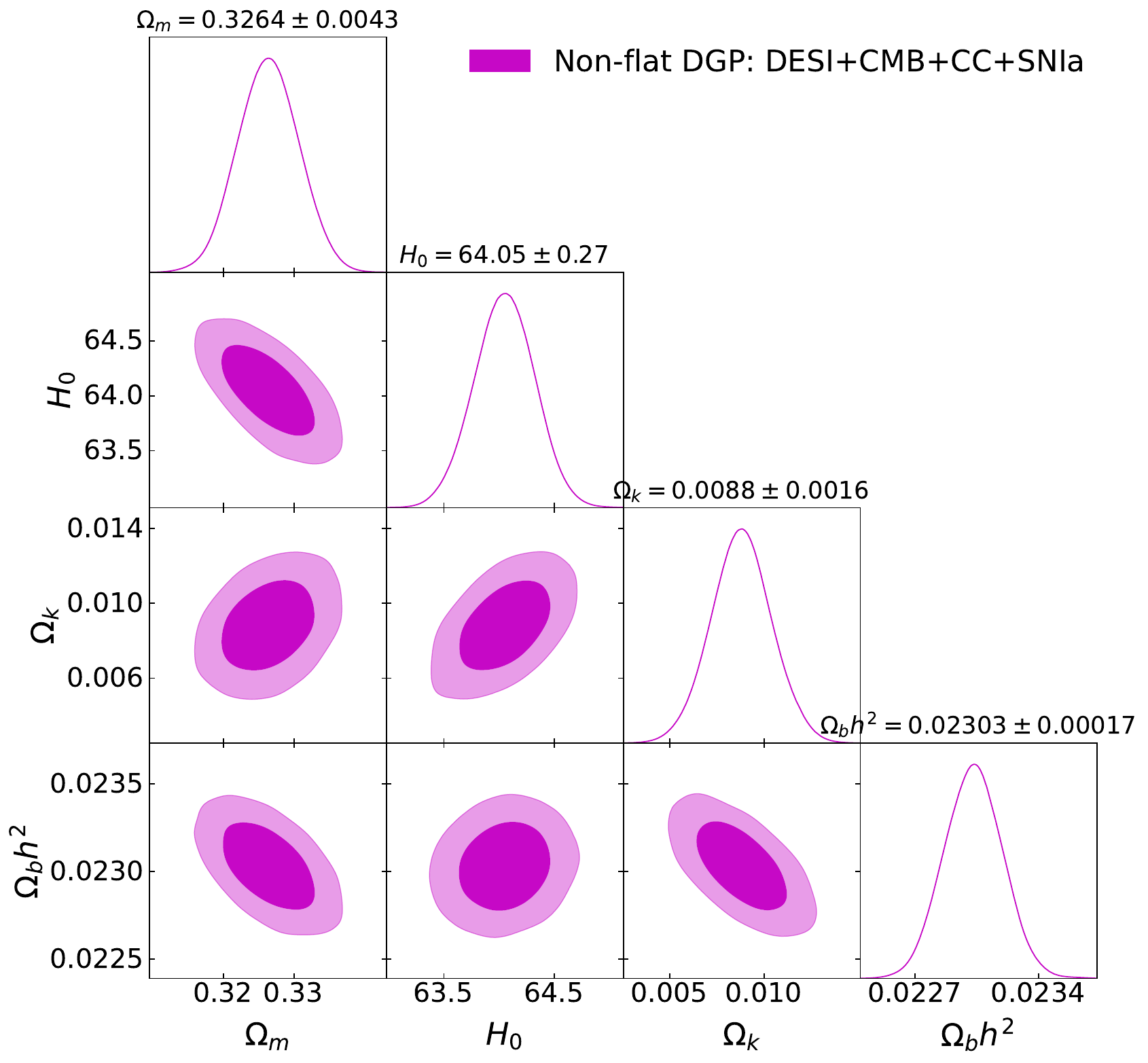}
    \caption{The one-dimensional marginalized probability distribution and two-dimensional contour plots of parameters for the non-flat DGP model at the 68\% and 95\% confidence levels,
    derived using DESI+CMB+CC+SNIa.}
    \label{fig:1}
\end{figure}

To evaluate and compare the statistical performance of the DGP model against the base $\mathrm{\Lambda}$CDM model, we employ the Akaike Information Criterion (AIC)~\cite{Akaike1974}, defined as:
\begin{equation}
\text{AIC} = \chi^2_{\min} + 2k,
\end{equation}
where $\chi^2_{\min}$ denotes the minimum chi-square and $k$ is the number of free parameters. The relative performance between the DGP and $\mathrm{\Lambda}$CDM models is quantified by $\Delta \text{AIC} = \text{AIC}_{\text{DGP}} - \text{AIC}_{\mathrm{\Lambda}CDM}$. 
According to information-theoretic interpretation, a difference of $\Delta \text{AIC} < 2$ indicates that the models are statistically indistinguishable, while $2 < \Delta \text{AIC} < 6$ suggests positive evidence in favor of the model with the lower AIC. Furthermore, $6 < \Delta \text{AIC} < 10$ indicates strong evidence in favor of the model with the lower AIC, while $\Delta \text{AIC} > 10$ implies that the model with the higher AIC is not supported by the observational data.

\section{Results and Discussion}
\label{sec:4}
\subsection{Cosmological Constraints and Model Comparison}
\begin{figure}[bp]
    \centering
    \includegraphics[width=\columnwidth]{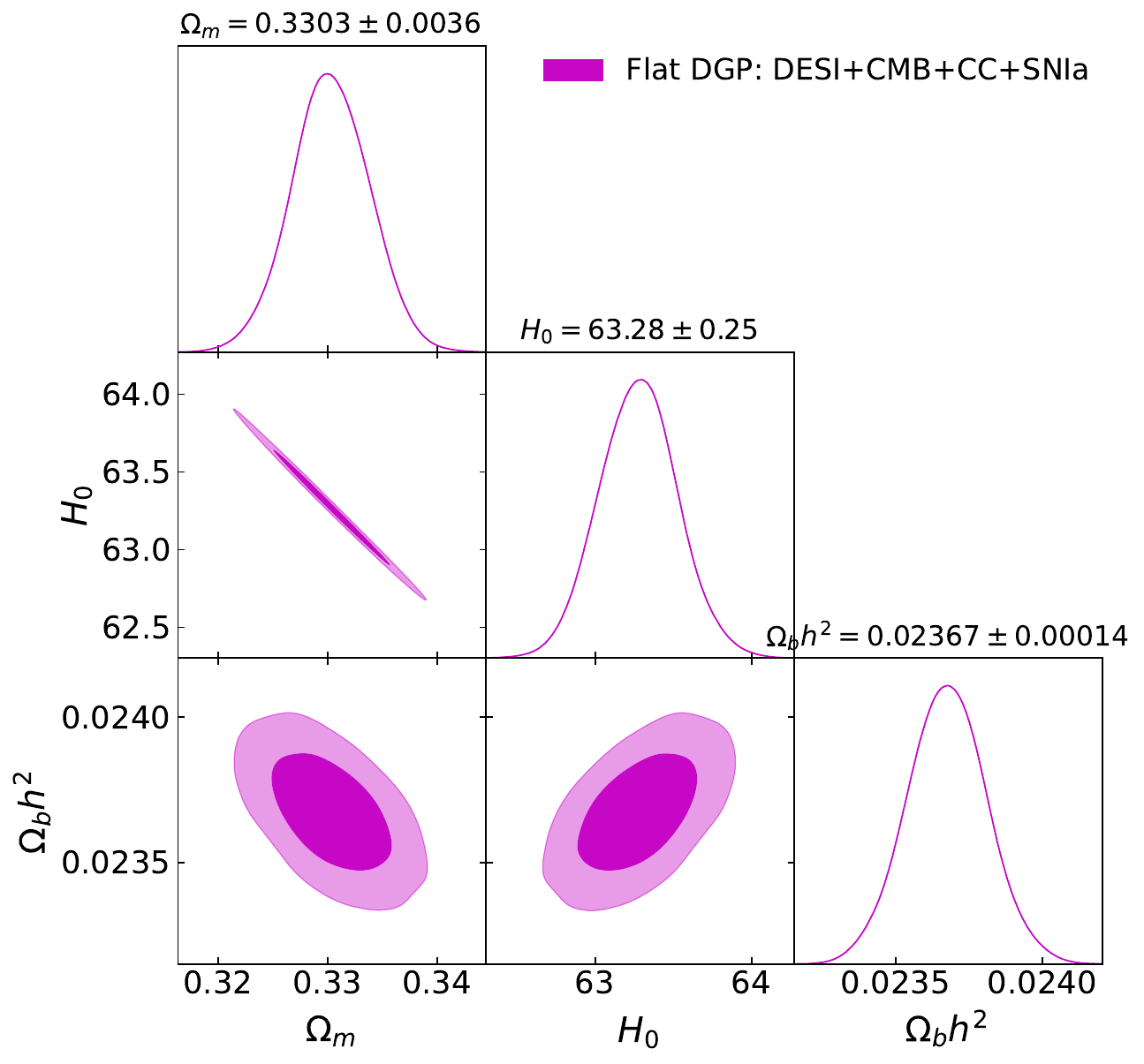}
    \caption{The one-dimensional marginalized probability distribution and two-dimensional contour plots of parameters for the flat DGP model at the 68\% and 95\% confidence levels,
    derived using DESI+CMB+CC+SNIa.}
    \label{fig:2}
\end{figure}

By leveraging the latest BAO measurements from DESI DR2 in combination with CC, SNIa, and CMB distance priors, we provide updated constraints on both flat and non-flat DGP models. The $1\sigma$ and $2\sigma$ confidence contours for these models are illustrated in Fig.~\ref{fig:1} and Fig.~\ref{fig:2}, respectively. As demonstrated in these figures, the joint analysis of the DESI+CMB+CC+SNIa datasets significantly tightens the parameter space, yielding more stringent constraints compared to individual probes.

Following a joint analysis using the combined DESI+CMB+CC+SNIa dataset, the marginalized constraints on the parameters for both flat and non-flat DGP models (at the $1\sigma$ confidence level) are summarized as follows:

For the non-flat DGP Model:

\begin{itemize}
    \renewcommand{\labelitemi}{$\bullet$}
    \item $\Omega_m = 0.3264 \pm 0.0043$,
    \item $H_0 = 64.05 \pm 0.27$,
    \item $\Omega_k = 0.0088 \pm 0.0016$,
    \item $\Omega_{b}h^{2} = 0.02303 \pm 0.00017$,
    \item $\Omega_{rc} = 0.1114 \pm 0.0016$.
\end{itemize}

For the flat DGP Model:

\begin{itemize}
    \renewcommand{\labelitemi}{$\bullet$}
    \item $\Omega_m = 0.3303 \pm 0.0036$,
    \item $H_0 = 63.28 \pm 0.25$,
    \item $\Omega_{b}h^{2} = 0.02367 \pm 0.00014$,
    \item $\Omega_{rc} = 0.1121 \pm 0.0013$.
\end{itemize}

\begin{table*}[htb]
\centering
\caption{The $\chi^2$ statistic values, total $\chi^2_{\text{min}}$, degrees of freedom ($dof$), AIC, and relative $\Delta\text{AIC}$ for each model in the DESI+CMB+CC+SN joint dataset. The $\Delta\text{AIC}$ for the DGP variants is calculated relative to their respective $\mathrm{\Lambda}$CDM baselines.}
\label{tab:chi2_stats}
\begin{small} 
\setlength{\tabcolsep}{6pt} 
\begin{tabular}{lcccccccc}
\toprule
Model & $\chi^2_{\text{DESI}}$ & $\chi^2_{\text{CC}}$ & $\chi^2_{\text{CMB}}$ & $\chi^2_{\text{SN}}$ & $\chi^2_{\text{min}}$ & $dof$ & AIC & $\Delta\text{AIC}$ \\
\hline
Non-flat $\mathrm{\Lambda}$CDM & 10.928 & 14.691 & 0.032 & 1036.812 & 1062.463 & 1088 & 1070.463 & --- \\
Non-flat DGP & 63.124 & 16.711 & 19.238 & 1063.211 & 1162.284 & 1086 & 1172.284 & +101.821 \\
\hline
Flat $\mathrm{\Lambda}$CDM & 10.947 & 14.709 & 0.027 & 1036.801 & 1062.484 & 1090 & 1068.484 & --- \\
Flat DGP & 66.724 & 16.021 & 72.501 & 1060.491 & 1215.737 & 1088 & 1223.737 & +155.253 \\
\bottomrule
\end{tabular}
\end{small}
\end{table*}

A comparison with the standard $\mathrm{\Lambda}$CDM model reveals that both flat and non-flat DGP scenarios prefer a significantly lower Hubble constant ($H_0 \sim 63$--$64 \mathrm{~km\,s^{-1}\,Mpc^{-1}}$) than the \textit{Planck} 2018 $\mathrm{\Lambda}$CDM value ($H_0 = 67.4 \pm 0.5 \mathrm{~km\,s^{-1}\,Mpc^{-1}}$~\cite{Planck:2018vyg}). As a result, the DGP model fails to alleviate the Hubble tension; instead, it exacerbates the discrepancy with local measurements. Allowing for spatial curvature leads to a mild preference for an open Universe, with $\Omega_k = 0.0088 \pm 0.0016$, corresponding to a $5.5\sigma$ deviation from spatial flatness. Meanwhile, the inferred values of $\Omega_m$ remain broadly consistent with $\mathrm{\Lambda}$CDM, suggesting that the primary deviations arise from the modified expansion history rather than the matter sector. Overall, introducing curvature produces only a marginal shift in parameters and does not significantly improve the consistency between early- and late-time observations.

To assess the goodness-of-fit, we compute the minimum $\chi^2$ and the Akaike Information Criterion (AIC) for each model, as summarized in Table~\ref{tab:chi2_stats}. 
Our statistical analysis demonstrably indicates that the $\mathrm{\Lambda}$CDM model provides a substantially better fit than the DGP scenarios.
In particular, the flat and non-flat DGP models are strongly disfavored, with large statistical penalties of $\Delta \text{AIC} = +155.253$ and $+101.821$, respectively. 
This discrepancy is primarily driven by the inability of the DGP model to simultaneously fit the CMB distance priors and DESI BAO measurements.
For example, in the flat case, the DGP model yields $\chi^2_{\rm CMB} = 72.501$ and $\chi^2_{\rm DESI} = 66.724$, which are dramatically larger than the corresponding $\mathrm{\Lambda}$CDM values ($0.027$ and $10.947$), indicating a strong tension with geometric constraints.

Allowing for spatial curvature partially alleviates this tension. 
In the non-flat DGP scenario, the total $\chi^2_{\rm min}$ decreases from $1215.737$ to $1162.284$, mainly due to a significant improvement in the CMB fit ($\chi^2_{\rm CMB} = 19.238$). 
However, despite this improvement, the non-flat DGP model remains heavily disfavored relative to the non-flat $\mathrm{\Lambda}$CDM baseline, with $\Delta \text{AIC} = +101.821$. 
This substantial excess in $\chi^2$ indicates that the DGP framework fails to simultaneously accommodate the combined constraints from DESI and \textit{Planck}, highlighting the limitations of its modified expansion history.

\subsection{Impact of the Sound Horizon Calibration}

To investigate the origin of potential internal tensions within the non-flat DGP framework, we perform a joint cosmological analysis by combining the latest DESI BAO measurements with CMB distance priors, CC, and SNIa datasets. 
Within this setup, we examine three distinct treatments of the sound horizon at the drag epoch, $r_d$. 
The resulting marginalized constraints on $H_0$ and $\Omega_k$, together with the corresponding AIC values, are summarized in Table~\ref{tab:2}.

It is important to note that the CMB distance priors are calibrated using the sound horizon at the photon decoupling epoch, $r_s$, whereas DESI BAO measurements probe the sound horizon at the drag epoch, $r_d$. 
These two quantities correspond to different physical epochs and are therefore not strictly identical. 
In a joint analysis combining CMB and BAO datasets, a consistent definition of the standard ruler is required. 
This is effectively achieved by enforcing $r_s \simeq r_d$ within the theoretical framework. 
Without such consistency, the mismatch between the two scales would artificially inflate the $\chi^2$ and bias the inferred cosmological parameters.

When $r_d$ is treated as a free parameter, effectively decoupled from early-universe physics, the inferred Hubble constant shifts to $H_0 = 73.57 \pm 1.10\,\mathrm{km\,s^{-1}\,Mpc^{-1}}$, in apparent agreement with local measurements. This behavior is accompanied by a substantial increase in spatial curvature, $\Omega_k = 0.0356 \pm 0.0022$, indicating that the model compensates for its modified expansion history by exploiting additional geometric freedom. The significantly lower AIC value in this case reflects an improved statistical fit; however, this improvement is achieved at the expense of physical consistency, as the standard ruler is no longer anchored to early-universe physics.
\begin{table}[tbp]
\centering
\caption{
Comparison of cosmological constraints for the non-flat DGP model under different sound horizon treatments.
Here $H_0$ is in units of km\,s$^{-1}$\,Mpc$^{-1}$, and $r_d$ is given in Mpc.
}
\label{tab:2}
\begin{small} 
\renewcommand{\arraystretch}{1.2} 
\setlength{\tabcolsep}{4pt}
\begin{tabular}{lccc}
\toprule
 & Fixed $r_d$ & Free $r_d$ & $r_d = r_s$ \\
\midrule
$H_0$ & $65.76 \pm 0.31$ & $73.57 \pm 1.10$ & $64.05 \pm 0.27$ \\
$\Omega_k$ & $0.0098 \pm 0.0012$ & $0.0356 \pm 0.0022$ & $0.0088 \pm 0.0016$ \\
$r_d$ & $147.09$ & $133 \pm 1.3$ & $148.1 \pm 0.4$ \\
AIC & $1215.985$ & $1103.283$ & $1172.284$ \\
\bottomrule
\end{tabular}
\end{small}
\end{table}

Fixing $r_d$ to the fiducial $\mathrm{\Lambda}$CDM value ($147.09$ Mpc) restores the early-time calibration, leading to a reduced Hubble constant of $H_0 = 65.76 \pm 0.31\,\mathrm{km\,s^{-1}\,Mpc^{-1}}$. The corresponding curvature remains moderate, $\Omega_k = 0.0098 \pm 0.0012$, but the overall fit quality deteriorates, as reflected by the increased AIC. This suggests that imposing a fixed standard ruler without fully accounting for the model-dependent early-universe physics introduces residual tension in the fit.

Adopting a fully self-consistent treatment ($r_d = r_s$), where the sound horizon is computed within the same cosmological framework, yields $H_0 = 64.05 \pm 0.27\,\mathrm{km\,s^{-1}\,Mpc^{-1}}$ and $\Omega_k = 0.0088 \pm 0.0016$. Compared to the free-$r_d$ case, the parameter uncertainties are significantly reduced, indicating that the physical linkage between early- and late-Universe quantities effectively breaks degeneracies. Although the AIC is higher than in the free-$r_d$ scenario, this result reflects a more physically meaningful description of the cosmological model.

These results indicate that the apparent agreement with local $H_0$ measurements in the free-$r_d$ case is driven by the relaxation of early-Universe constraints rather than a genuine improvement of the model. Once a consistent calibration of the sound horizon is imposed, the inferred value of $H_0$ shifts toward lower values, and the tension with local measurements reappears. This highlights the importance of maintaining physical consistency in joint cosmological analyses.

This motivates the adoption of the self-consistent treatment ($r_d = r_s$) as the baseline choice in the following analysis.

\subsection{Impact of Dataset Combinations on Parameter Constraints}

\begin{figure}[bh]
    \centering
    \includegraphics[width=\columnwidth]{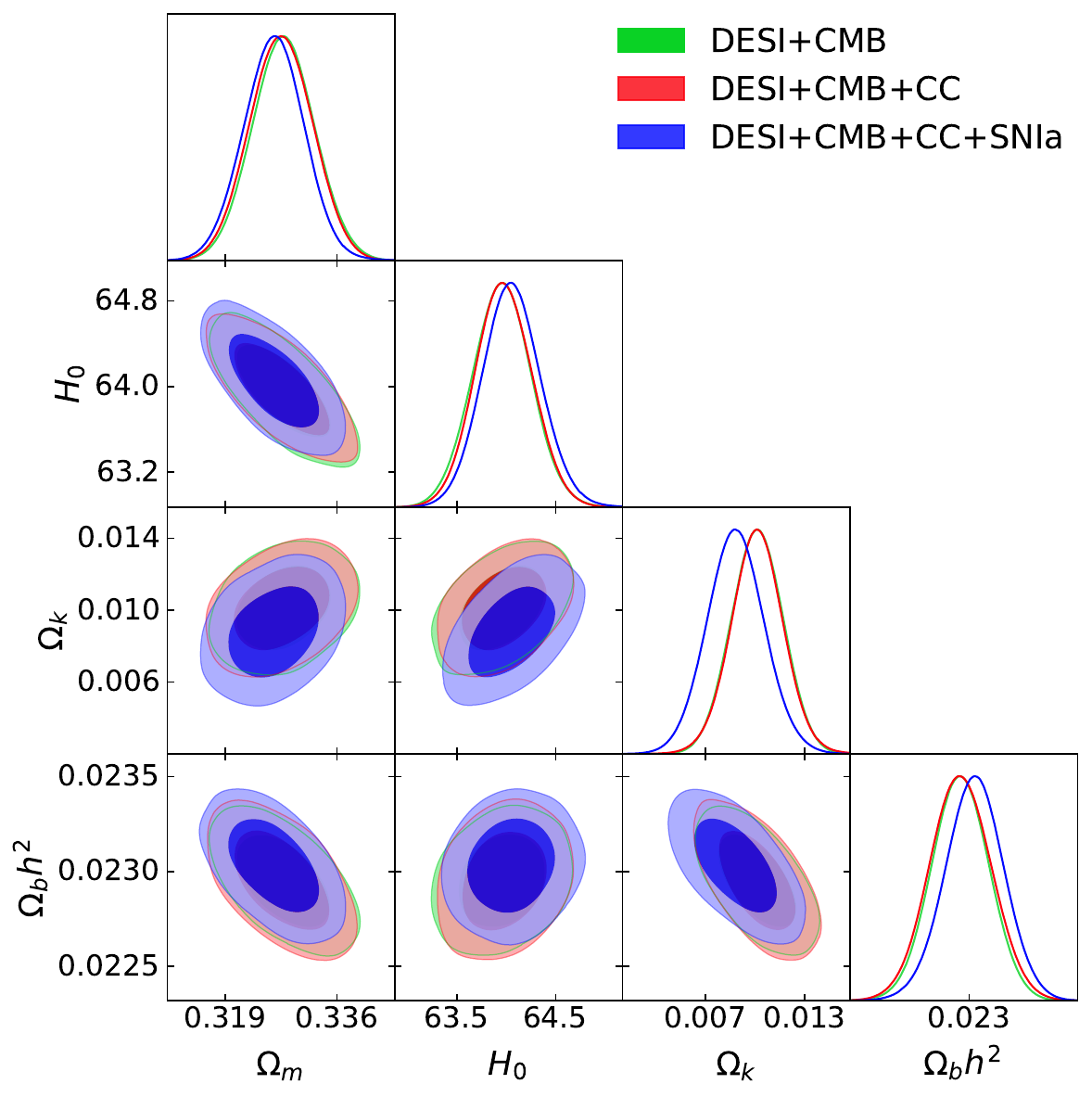}
    \caption{The one-dimensional marginalized probability distributions and two-dimensional $1\sigma$ and $2\sigma$ confidence contours for the non-flat DGP model. Three dataset combinations are compared: DESI+CMB (green), DESI+CMB+CC (red), and the full joint dataset DESI+CMB+CC+SNIa (blue). }
    \label{fig:3}
\end{figure}

\begin{table*}[bt]
\centering
\caption{
Comparison of cosmological parameter constraints for the non-flat DGP model under successive dataset combinations. The quoted values correspond to the $1\sigma$ confidence level.(Here $H_0$ is in units of km\,s$^{-1}$\,Mpc$^{-1}$.)
}
\label{tab:3}
\begin{footnotesize} 
\setlength{\tabcolsep}{3pt}
\begin{tabular}{@{}lcccccc@{}} 
\bottomrule
Datasets & Model & $\Omega_m$ & $\Omega_k$ & $H_0$ & $\Omega_b h^2$ & $\Omega_{r_c}$\\
\hline
\textbf{DESI+CMB} & $\mathrm{\Lambda}$CDM & $0.2973 \pm 0.0042$ & $0.0002 \pm 0.0012$ & $69.01 \pm 0.32$ & $0.02258 \pm 0.00016$ & --- \\
                  & DGP          & $0.3280 \pm 0.0046$ & $0.0101 \pm 0.0015$ & $63.95 \pm 0.29$ & $0.02295 \pm 0.00016$ & $0.1106 \pm 0.0017$ \\
\hline
\textbf{DESI+CMB+CC} & $\mathrm{\Lambda}$CDM & $0.2979 \pm 0.0041$ & $0.0003 \pm 0.0012$ & $68.94 \pm 0.31$ & $0.02256 \pm 0.00016$ & --- \\
                     & DGP          & $0.3275 \pm 0.0045$ & $0.0100 \pm 0.0015$ & $63.97 \pm 0.27$ & $0.02296 \pm 0.00016$ & $0.1108 \pm 0.0016$ \\
\hline
\textbf{DESI+CMB+CC+SN} & $\mathrm{\Lambda}$CDM & $0.2985 \pm 0.0040$ & $0.0001 \pm 0.0013$ & $69.07 \pm 0.31$ & $0.02259 \pm 0.00016$ & --- \\
                        & DGP          & $0.3264 \pm 0.0043$ & $0.0088 \pm 0.0016$ & $64.05 \pm 0.27$ & $0.02303 \pm 0.00017$ & $0.1114 \pm 0.0016$ \\
\bottomrule
\end{tabular}
\end{footnotesize}
\end{table*}

Building on the self-consistent treatment of the sound horizon ($r_d = r_s$), we investigate how successive dataset combinations affect the parameter constraints in the non-flat DGP framework. 
We consider three progressively extended datasets: DESI BAO combined with CMB distance priors (DESI+CMB), with the addition of cosmic chronometers (DESI+CMB+CC), and the full joint dataset including Type Ia supernovae (DESI+CMB+CC+SNIa). 
The corresponding constraints are illustrated in Fig.~\ref{fig:3} and summarized in Table~\ref{tab:3}.

A clear contrast emerges between the $\mathrm{\Lambda}$CDM and DGP frameworks as additional observational probes are included. 
For the $\mathrm{\Lambda}$CDM model, the inferred cosmological parameters remain remarkably stable across all dataset combinations. 
In particular, $\Omega_m$, $\Omega_k$, and $H_0$ exhibit only minimal shifts as CC and SNIa data are added to the DESI+CMB baseline, indicating a high level of internal consistency among different cosmological probes.

In contrast, the DGP model shows a noticeably different behavior. 
While the inclusion of additional datasets slightly tightens the parameter uncertainties, the central values exhibit systematic shifts, particularly in $\Omega_m$, $\Omega_k$, and $H_0$. 
Notably, the DGP model consistently favors a higher matter density ($\Omega_m \sim 0.326$--0.328) and a lower Hubble constant ($H_0 \sim 63.9$--64.1), compared to the $\mathrm{\Lambda}$CDM results.

Moreover, the preference for a non-zero spatial curvature in the DGP scenario persists across all dataset combinations, with $\Omega_k \sim 0.009$--0.010, indicating a stable deviation from spatial flatness at the level of several standard deviations. 
This behavior suggests that the DGP framework requires a non-negligible curvature contribution to partially compensate for its modified expansion dynamics.

Taken together, while the $\mathrm{\Lambda}$CDM model demonstrates strong robustness under the addition of independent datasets, the DGP model exhibits mild parameter shifts and persistent tension, reflecting its limited ability to simultaneously accommodate multiple cosmological probes within a consistent framework.

\subsection{Evolution of the Deceleration Parameter}

\begin{figure}[htb]
    \centering
    \includegraphics[width=\columnwidth]{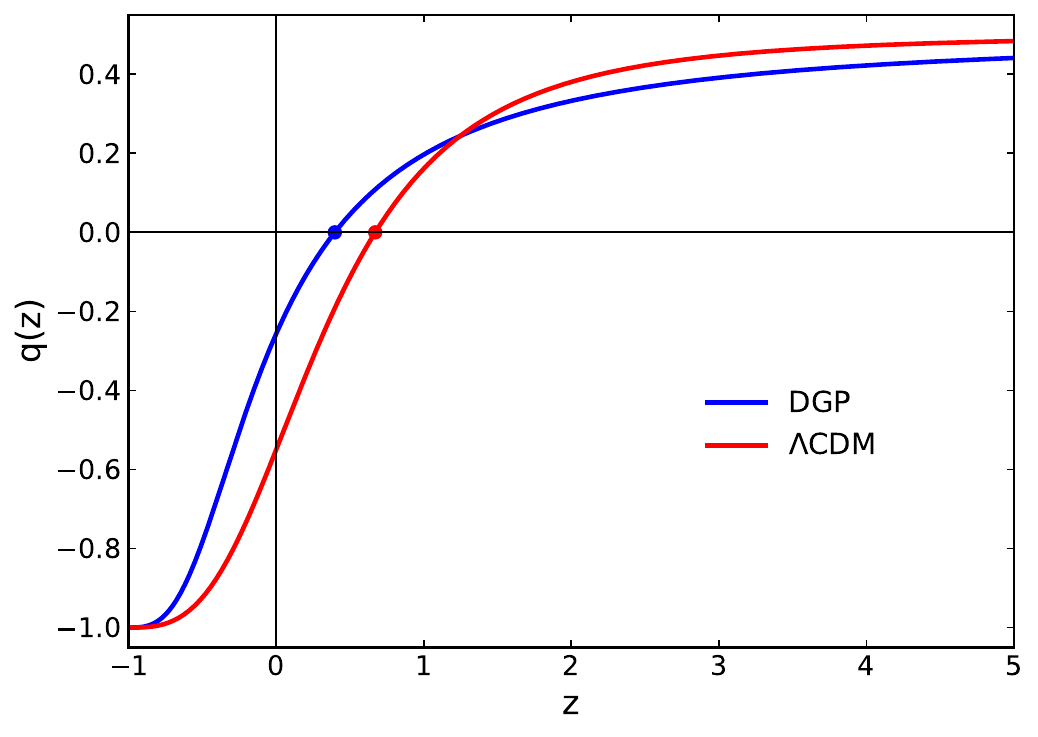}
    \caption{Evolution of the deceleration parameter $q(z)$ as a function of redshift for the non-flat DGP and $\mathrm{\Lambda}$CDM models, derived from the DESI+CMB+CC+SNIa dataset. The transition from deceleration ($q>0$) to acceleration ($q<0$) is clearly shown, with the DGP model exhibiting a delayed onset of cosmic acceleration compared to $\mathrm{\Lambda}$CDM.}
    \label{fig:4}
\end{figure}

To further investigate the dynamical evolution of the Universe within the DGP framework, we reconstruct the redshift dependence of the deceleration parameter $q(z)$ for the non-flat DGP and $\mathrm{\Lambda}$CDM models, using the full joint dataset DESI+CMB+CC+SNIa under the self-consistent sound horizon treatment ($r_d = r_s$). The results are shown in Fig.~\ref{fig:4}. 

The evolution of $q(z)$ exhibits the expected transition from an early decelerating phase ($q>0$) to a late-time accelerating phase ($q<0$), indicating that the model can qualitatively reproduce the observed cosmic acceleration. However, clear quantitative deviations from the standard $\mathrm{\Lambda}$CDM model are present.

In particular, the non-flat DGP model predicts systematically larger values of $q(z)$ over the entire redshift range, especially at low redshifts, implying a weaker acceleration compared to $\mathrm{\Lambda}$CDM. This behavior is further reflected in the transition redshift, which is found to be $z_t^{\rm non\text{-}flat} = 0.41$, significantly lower than $z_t^{\mathrm{\Lambda}{\rm CDM}} \approx 0.67$.

This delayed onset of cosmic acceleration indicates that the DGP model is less efficient in driving late-time acceleration. Such behavior is fully consistent with the parameter constraints obtained in previous sections, where the model favors a lower $H_0$ and shows significant tension with CMB and BAO observations.

Taken together, these results provide dynamical evidence that the DGP framework cannot simultaneously reproduce the expansion history required by both early- and late-Universe probes. This reinforces the conclusions drawn from the $\chi^2$ and AIC analyses, highlighting the intrinsic limitations of the standard DGP model in explaining current cosmological observations.

\subsection{Comparison with Previous Constraints}

\begin{table*}[htb]
    \caption{Cosmological parameter constraints from various studies with $1\sigma$ uncertainties.}
    \label{tab:4}
    \begin{center}
    \centering
    \footnotesize
    \setlength{\tabcolsep}{3pt}
    \begin{tabular}{lcccccc}
    \toprule
    Data set & Model & $H_0$ & $\Omega_{m}$ & $\Omega_{r_c}$ & $\Omega_{k}$  & Ref. \\
      &   & [$\mathrm{km\,s^{-1}\,Mpc^{-1}}$] &   &   &   &   \\
    \midrule
    Gold+SNLS+SDSS & \textbf{Non-flat} & -- & $0.27^{+0.018}_{-0.017}$ & $0.216^{+0.012}_{-0.013}$ & $-0.350^{+0.080}_{-0.083}$ &~\cite{Guo:2006ce} \\
    \midrule
    \makecell[l]{SNIa+GRBs+CMB\\+BAO+CBF+CC} & \textbf{Non-flat} & -- & $0.235^{+0.125}_{-0.074} $ & $0.138^{+0.031}_{-0.036}$ & $0.033$ &~\cite{Liang_2011} \\
    \midrule
    \makecell[l]{SNLS+CMB+SDSS\\+LSS+gas} & \textbf{Non-flat} & -- & $0.21\pm0.01$ & $0.16\pm0.01$ & $0.01\pm0.04$  &~\cite{Movahed:2007ie} \\
    \midrule
    lens+CMB+SDSS & Flat & -- & $0.32^{+0.02}_{-0.02}$ & $0.116\pm0.007$ & --  &~\cite{PhysRevD.84.023005} \\
    \midrule
    13 lens & Flat & -- & $0.30^{+0.19}_{-0.11}$ & $0.123^{+0.057}_{-0.042}$ & -- &~\cite{Zhu:2008sg} \\
    \midrule
    63 lens & Flat & -- & $0.22^{+0.10}_{-0.09}$  & $0.152^{+0.0365}_{-0.0371}$ & -- &~\cite{Cao_2012} \\
    \midrule
    $6D_{\Delta t}+4D_d+QSO$ & \textbf{Non-flat} & $66.35^{+14.92}_{-8.10}$ & $0.253^{+0.086}_{-0.083}$ & $0.150^{+0.110}_{-0.104}$ & $-0.047^{+0.384}_{-0.311}$ &~\cite{Liu:2022mpj} \\
      & Flat & $67.84^{+1.10}_{-1.12}$ & $0.246^{+0.043}_{-0.037}$ & $0.142^{+0.014}_{-0.016}$ & -- &~\cite{Liu:2022mpj} \\
    \midrule
    \makecell[l]{SNIa+BAO+CMB+\\GRBs+CBF+LT+GF} & \textbf{Non-flat} & $66.798^{+2.85}_{-2.483}$ & $0.270^{+0.0278}_{-0.0323}$ & -- & $0.0123^{+0.00789}_{-0.00993}$ &~\cite{Xu:2010um} \\
     & Flat & $65.219^{+2.190}_{-1.956}$ & $0.265^{+0.0291}_{-0.0302}$ & -- & -- &~\cite{Xu:2010um}  \\
    \midrule
    DESI+CMB+CC+SNIa  & \textbf{Non-flat} & $ 64.05 \pm 0.27 $ & $ 0.3264\pm 0.0043 $ & $ 0.1114 \pm 0.0016 $ & $0.0088\pm 0.0016$ &This work\\
     &Flat & $ 63.28 \pm 0.25 $ & $ 0.3303 \pm 0.0036 $ & $ 0.1121 \pm 0.0013 $ & --  &This work\\

    \midrule
    \end{tabular}
    \end{center}
    \footnotesize{
    \textbf{Abbreviations:} SNIa = Type Ia Supernovae; BAO = Baryon Acoustic Oscillations; CC = Cosmic Chronometers; CMB = Cosmic Microwave; CBF = Cluster Baryon Fraction Background; GF = Galaxy Cluster Gas Fraction; GRBs = Gamma-Ray Bursts; lens = gravitational lensing surveys; LSS = Large-Scale Structure; LT = Lookback Time; QSO = Quasars; SDSS = Sloan Digital Sky Survey; $D_{\Delta t}$ = the time-delay distance; $D_d$ = the angular diameter distance to the lens.}

\end{table*}

Table~\ref{tab:4} summarizes cosmological parameter constraints reported in previous studies, together with the results obtained in this work, all quoted at the $1\sigma$ confidence level. 
The table compares both non-flat and flat cosmological models across multiple observational probes, including Type Ia supernovae (e.g., from the Gold, SNLS, and SDSS datasets), CMB, BAO, CC, Gamma-Ray Bursts (GRBs), Cosmic Background Fields (CBF), Large-Scale Structure (LSS), gravitational lensing surveys, etc. Through systematic analysis of the constraint results from different datasets in Table~\ref{tab:4}, we observe significant improvements in the precision of cosmological parameter measurements. The following is a comparative analysis of the results from different research:

\begin{itemize}
    \renewcommand{\labelitemi}{$\bullet$}
    \item Ref.~\cite{Liu:2022mpj} analyzed 6 time-delay lenses ($D_{\Delta t}$) and 4 angular diameter distances ($D_d$) combined with QSO samples, measuring $H_0 = 66.35^{+14.92}_{-8.10} \, \text{km s}^{-1} \text{Mpc}^{-1}$ (non-flat) with a relative uncertainty of about 20\%. Ref.~\cite{Xu:2010um} used SN+BAO+CMB+GRBs+CBF+LT+GF data and obtained $H_0 = 66.798^{+2.85}_{-2.483} \, \text{km s}^{-1} \text{Mpc}^{-1}$ for a non-flat universe. 
    In comparison, our study based on the DESI DR2 joint dataset shows $H_0 = 64.05 \pm 0.27 \, \text{km s}^{-1} \text{Mpc}^{-1}$ (non-flat), significantly reducing the error range by roughly an order of magnitude compared to previous studies and achieving sub-percent level precision ($0.4\%$).

    \item For the non-flat DGP model, varying cosmic curvatures will lead to different values of $\Omega_m$. Therefore, we compare and analyze the restricted $\Omega_m$ results for the flat DGP here. Ref.~\cite{Xu:2010um} gives $\Omega_m = 0.265^{+0.0291}_{-0.0302}$. 
    Our updated result for the flat DGP model yields a notably higher matter density, $\Omega_m = 0.3303 \pm 0.0036$. While the central value has shifted to accommodate the modern CMB and BAO calibrations, the constraint precision is improved by a factor of nearly $8$ compared to Ref.~\cite{Xu:2010um}.

    \item The curvature parameter $\Omega_k$ exhibits distinct results across different studies. Ref.~\cite{Guo:2006ce}, relying on the early Gold+SNLS+SDSS data, measured $\Omega_k$ to show a strong negative curvature with a specific value of $-0.350^{+0.080}_{-0.083}$. Ref.~\cite{Xu:2010um} presented a result of $\Omega_k = 0.0123^{+0.00789}_{-0.00993}$. 
    his is highly consistent with our updated research result of $\Omega_k = 0.0088 \pm 0.0016$, indicating a stable and significant preference for a slightly open universe (positive curvature).

    \item Most remarkably, the DESI DR2 joint data improves the precision of the crossover scale parameter $\Omega_{r_{c}}$ to $0.1121 \pm 0.0013$ (flat), achieving an accuracy approximately $5$ times better than the lens+CMB+SDSS result ($0.116 \pm 0.007$) reported in Ref.~\cite{PhysRevD.84.023005}.
\end{itemize}

Modern multi-messenger astronomy has significantly advanced cosmological parameter measurements, as demonstrated by our comparative analysis of DGP models across different datasets. While Ref.~\cite{Guo:2006ce} and~\cite{Liang_2011} report transition redshifts of $z_t = 0.86^{+0.07}_{-0.08}$ and $0.67^{+0.03}_{-0.04}$ respectively, 
our updated analysis yields values of $z_t = 0.41^{+0.0126}_{-0.0125}$ (non-flat) and $z_t = 0.40^{+0.0110}_{-0.0107}$ (flat). 
These results indicate that the transition redshift is only weakly sensitive to spatial curvature, suggesting that the delayed onset of cosmic acceleration is an intrinsic feature of the DGP framework rather than being driven by geometric effects. 
These findings highlight both similarities and differences in parameter estimation across studies, and suggest that the modified expansion history in DGP models may require further refinement at the level of braneworld gravity.
\section{Conclusions}
\label{sec:5}

In this work, we have presented updated and stringent constraints on both flat and non-flat DGP models by combining the latest DESI DR2 BAO measurements with CC, SNIa, and CMB distance priors within a consistent framework. 
For the non-flat DGP model, we obtain $H_0 = 64.05 \pm 0.27\, \mathrm{km\,s^{-1}\,Mpc^{-1}}$, $\Omega_m = 0.3264 \pm 0.0043$, and $\Omega_k = 0.0088 \pm 0.0016$, indicating a preference for a mildly open Universe at the $\sim 5.5\sigma$ level. 
For the flat DGP model, the constraints are $H_0 = 63.28 \pm 0.25\, \mathrm{km\,s^{-1}\,Mpc^{-1}}$, $\Omega_m = 0.3303 \pm 0.0036$, and $\Omega_{r_c} = 0.1121 \pm 0.0013$. 
In both cases, the inferred Hubble constant is significantly lower than the \textit{Planck} $\mathrm{\Lambda}$CDM value, indicating that the DGP framework exacerbates rather than alleviates the Hubble tension.

From the perspective of model comparison, the DGP scenario is strongly disfavored by current observations. 
The large excess in $\chi^2$ and the corresponding $\Delta \mathrm{AIC}$ values ($\sim 100$--150) show that both flat and non-flat DGP models fail to provide a competitive fit relative to the standard $\mathrm{\Lambda}$CDM paradigm. 
This discrepancy is mainly driven by the difficulty of the DGP model in simultaneously accommodating the geometric constraints from DESI BAO and the CMB distance priors.
The dynamical reconstruction further supports this conclusion. 
The transition redshift is found to be $z_t \simeq 0.41$, significantly lower than the $\mathrm{\Lambda}$CDM expectation ($z_t \sim 0.67$), indicating a delayed onset of cosmic acceleration, since a smaller value of $z_t$ corresponds to a later transition from deceleration to acceleration.
This reflects a weaker late-time acceleration, consistent with the systematically higher values of $q(z)$ and the lower inferred $H_0$. 
Compared with previous studies~\cite{Guo:2006ce,Liang_2011}, which reported larger transition redshifts, the present results highlight the impact of improved observational precision and updated datasets.

The role of sound horizon calibration is crucial in this context. 
Although treating $r_d$ as a free parameter can bring $H_0$ into apparent agreement with local measurements, it simultaneously leads to unphysical curvature values, indicating that such a resolution is not physically meaningful. 
Once a self-consistent treatment is imposed, the model is driven back toward lower $H_0$, and the tension persists.
Taken together, these results indicate that the standard DGP framework is unable to simultaneously satisfy early- and late-Universe observational constraints. 
Rather than resolving existing tensions, the model introduces additional inconsistencies when confronted with high-precision cosmological data. 
In this sense, the present work provides an updated and systematic reassessment of the DGP scenario using the latest DESI observations, while also consolidating and refining conclusions drawn in earlier studies.

The significantly improved precision provided by DESI highlights the growing constraining power of next-generation cosmological surveys. Future observations, including gamma-ray bursts (GRBs)~\cite{Dainotti:2020azn,Ryan:2019fhz}, fast radio bursts (FRBs)~\cite{Zhang:2022uzl,Bochenek:2020zxn}, the tip of the red giant branch (TRGB)~\cite{Freedman:2024eph}, and weak lensing measurements from the Euclid mission~\cite{Euclid:2024guw,Scognamiglio:2025qgn},
as well as growth-based probes such as redshift-space distortions ($f\sigma_8$)~\cite{eBOSS:2020yzd,DESI:2024jxi}, will provide crucial tests for the DGP model. Since modified gravity in the DGP framework inherently predicts a suppressed growth rate due to the leakage of gravity into extra dimensions, the inclusion of these observables is expected to break the degeneracy between geometry and growth, likely imposing even more stringent constraints and further quantifying the statistical tension with the $\mathrm{\Lambda}$CDM baseline,
will provide further opportunities to test modified gravity scenarios and explore possible extensions beyond the standard cosmological model.

\section{Acknowledgements}
We thank Fengquan Wu(NAOC), Xin Zhang(NEU), Yan Gong(NAOC), Lei Feng(PMO), Shuangxi Yi(QFNU), Yankun Qu(QFNU) and Fayin Wang(NJU) for their helpful suggestions and comments. 
This work is supported by the Shandong Provincial Natural Science Foundation (Grant
No. ZR2025MS16).

\newcommand{\bibcommenthead}{}
\bibliographystyle{sn-aps}
\bibliography{ref}

\end{document}